**Possible dimensionality transition behavior in localized plasmon resonances of confinement-controlled graphene devices**

Takahiro Morimoto*, Yoshiue Ikuta and Toshiya Okazaki

*CNT-Application Research Center, National Institute of Advanced Industrial Science and Technology (AIST), Tsukuba 305-8565, Japan*

E-mail: t-morimoto@aist.go.jp

**Abstract**

We investigated the dimensionality transition behavior of graphene localized plasmon resonances in confinement-controlled graphene devices. We first demonstrated a possibility of dimensionality transition, based on the devices' carrier-density dependence, from a two-dimensional plasmon resonance to a one-dimensional plasmon one. We fabricated optical transparent devices and electrical transport devices on the same optical transparent wafer. These devices allow detailed control and analysis between carrier density and plasmon resonance peak positions. The carrier density from $\sqrt[4]{n}$ (two-dimensional) to constant (one-dimensional) is consistent with the theoretical predictions based on the Dirac Fermion carriers in linear-band structure materials.

**Introduction**

The optical properties of low-dimensional carbon nanomaterials are attractive due to their unique behaviors in various frequency regions[1]. These materials show unique plasmonic behavior from far-infrared (FIR) to terahertz (THz) regions due to strongly confined-carrier dynamics[2–11]. Because of their high sensitivity to charge variations, surface modifications, and molecular adsorptions, plasmonic behavior is also important for highly efficient and sensitive optical components and detection devices[12]. Such behavior is also unique in dimensionality and carrier-density dependent[13]. It has recently been reported that the two-dimensionally defined graphene plasmons show weaker dependence on carrier density as $\sqrt[4]{n}$, not $\sqrt[2]{n}$ [7,10]. This behavior originated from truly mass-less Dirac Fermion natures in graphene devices. In the theoretical model, a graphene plasmon should show the disappearance of carrier density derivations in the one-dimensional plasmon resonance. However, there has not yet been any experimental demonstration of one-dimensional plasmon resonance in well-defined samples.

We present the dimensionality dependence originating from the confinement differences defined by the channel width of resonant cavities. These low-dimensional graphene plasmons showed clear transition from two dimensions to one dimension. The vanishing carrier-density dependence of plasmon peak positions are consistent with the theoretical predictions based on the mass-less Dirac fermion descriptions. These results are important for the fundamental understanding of low-dimensional localized plasmons. Moreover, these carrier density robustness are also important for fabricating controllable plasmon devices based on low-dimensional materials.

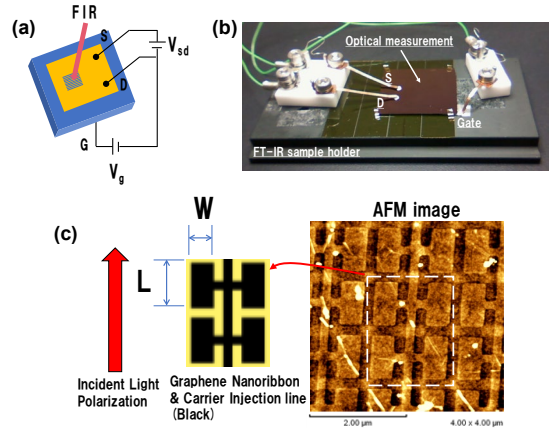

Figure 1. (color online)
(a) Schematic of measurement setup. FIR spectrum was measured at patterned area, and electric transport measurement was conducted at non-lithographic area. Gate-voltages were applied using back-gate configurations. (b) Optical image of sample holder used in this study. Samples were mounted on FT-IR sample holder. Electric contacts were done using hand-crafted probing systems. (c) Schematic of sample pattern and AFM images obtained after developed devices. Cavities of optical resonances were same length ($L$) but with variable widths ($W$) and connected to carrier injection lines.

Figure. 1(a) shows a schematic picture of our device configuration. The single-layer graphene was transferred on a $SiO_2$ (285 nm)/Si substrate. The conductivity of the Si substrate was carefully set to 10 Ωcm for FIR transparent measurement and back-gate configuration transport measurement on all device. The carrier density at each gate voltage was estimated from two parameters, Drude absorption in the FIR region and DC-conductance difference from the Dirac point at a conductance minimum in each device. Figure. 1(b) shows the hand-crafted measurement sample holder for this study. The graphene devices were mounted on the Fourier transform infrared spectroscopy (FT-IR) sample holder with three electrodes for electric transport measurements with conventional back-gate configurations. These devices were set into a vacuumed FT-IR system (Bruker: Vertex 80v) for reducing the gas-adsorption effect on the transport properties. Figure 1(c) shows atomic force microscopy (AFM) images of one of our devices with 673-nm-wide and 1.0-m-long plasmon cavities. These devices were fabricated using electron beam lithography and conventional lift-off process. We used the same resonance length of 1.0 μm but varied the widths from 186 to 990 nm. The resonance cavities were connected to the carrier injection line to change the carrier density depending on the gate voltage. The incident light polarization was parallel to the cavity-length directions. Therefore, the plasmon resonance should also be excited in the length direction.

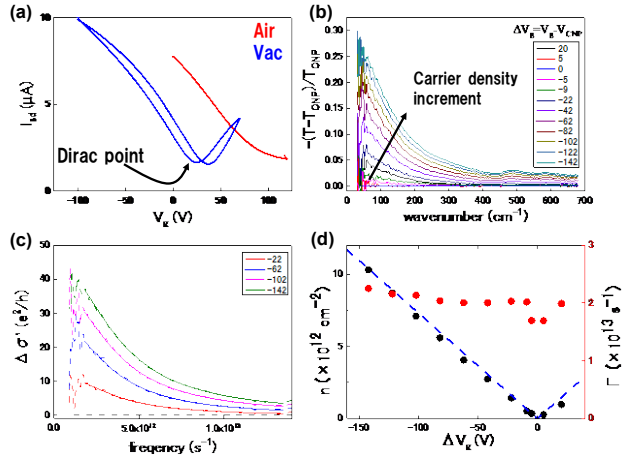

Figure 2. (color online)
(a) Source-drain current as function of back-gate voltage. In air, Dirac point was not observed due to shift by molecular adsorption (red curve). Blue curves indicate transport properties under vacuum. (b) Drude absorption in FIR region observed with non-patterned region at each gate voltage. By increasing back-gate voltages, Drude absorption monotonically increased by increasing carrier densities. (c) Carrier densities and scattering-time constant estimated from fitting of Drude absorptions. Carrier densities increased by applying gate voltages, and scattering-time constant was almost constant due to electrically carrier modulations.

Figure 2(a) shows the electrical transport properties depending on the back-gate voltage ($V_g$) at room temperature. In air, the charge neutral point (CNP) shifted to a positive higher gate voltage due to gas adsorptions, such as oxygen, onto the graphene surface. The blue curves show the transport properties under vacuum. These curves showed a clear minimum point ($V_{CNP}$) of the source-drain current (maximum resistance) corresponding to the Dirac point in our devices. Figure. 2(b) shows the normalized transmittance corrected by the spectra at $V_{CNP}$ ($T_{CNP}$). In these spectra, the Drude absorption behaviors were clearly observed in both gate-voltage directions with non-patterned regions. Each gate voltage was corrected as effective gate voltages ($\Delta V g$) estimated from the differences between applied gate voltages and $V_{CNP}$.

For precise estimation of the carrier density at each gate voltage from these optical absorptions, Drude behaviors are considered as optical responses due to the electric field modified by environmental correction factors[14,15]. The relationship between surface conductivity and optical transmittance is given by

$$-(T - T_{CNP})/T_{CNP}(\omega) = 4\pi/c \cdot Re\{\Delta\sigma(\omega) \times L\}, \tag{1}$$

where $(T - T_{CNP})/T_{CNP}$ is normalized transmittance, as mentioned above, $c$ is the speed of light, $\Delta\sigma$ is induced conductivity change due to the gate voltage, and $L$ is local field factors depending on the device configuration. The complex surface conductivity and local field factor are described as $\Delta\sigma = \Delta\sigma' + i\Delta\sigma''$ and $L = L' + iL''$, respectively. Our graphene devices were fabricated on the $SiO_2$/Si substrate with a 285-nm insulating layer. Therefore, the electric field near the devices should be modified, and this local field factor is given by

$$L = 1 + r_{ag} + t_{ag}r_{gs}t_{ga}\exp(2ikn_{SiO_2}d)/(1 - \exp(2ikn_{SiO_2}d)r_{gs}r_{ga}), \tag{2}$$

where $d$ is the thickness of the $SiO_2$ layer, $k$ is the wave number of irradiated light, $n_{SiO_2}$ is the complex refractive index of $SiO_2$, and $r$ and $t$ represent the Fresnel reflection coefficient and transmission coefficient, respectively. In the above equation, the subscripts correspond to air (a), glass (g), and silicon (s). For example, $r_{ag}$ is described as $(n_{air} - n_{SiO_2})/(n_{air} + n_{SiO_2})$. The relationship between normalized transmittance and surface conductivity is given by

$$-(T - T_{CNP})/T_{CNP}(\omega) = 4\pi/c \cdot Re\{\Delta\sigma(\omega) \times L\} = 4\pi/c \cdot (\Delta\sigma'(\omega)L' - \Delta\sigma''(\omega)L''). \quad (3)$$

In the FIR regions, the imaginary part of each material used in this study is negligibly small. Therefore, the last term in the above equation is also negligibly small in this measurement. Finally, the real part of surface conductivity is estimated as

$$\Delta\sigma'(\omega) = c/4\pi L' \cdot \{-(T - T_{CNP})/T_{CNP}(\omega)\}, \quad (4)$$

where $L'$ is 0.46 for the FIR region in our device configurations calculated from Eq. (2). Figure 2(c) shows the real part of surface-conductivity difference as a function of frequency ($s^{-1}$). The surface conductivity is clear due to Drude absorption phenomena. Consequently, we can describe surface conductivity as the following Drude absorption formula

$$\Delta\sigma'(\omega) = Re\{\sigma - \sigma_{CNP}\} = Re\{iD/\pi(\omega + i\Gamma) - iD_{CNP}/\pi(\omega + i\Gamma_{CNP})\}, \quad (5)$$

where $D$ is the Drude weight and $\Gamma$ is the scattering rate. After unit conversion from angular frequency to frequency, the above equation is simply described as

$$\Delta\sigma'(f) = D\Gamma/\pi(4\pi^2 f^2 + \Gamma^2) - D_{CNP}\Gamma_{CNP}/\pi(4\pi^2 f^2 + \Gamma^2) + C, \quad (6)$$

where C is a constant and is added for background correction. In Fig. 2(c), the fitting curves are indicated as solid curves at each $\Delta Vg$. These fitting curves showed good agreement with the surface conductivity curves. We used $D_{CNP} = 6.13 \times 10^{14} \, e^2/h \cdot s^{-1}$ and $\Gamma_{CNP} = 2\pi c \times 10^2 \, s^{-1}$. This scattering rate was commonly reported for transferred CVD-growth graphene devices[14,16,17]. The relationship between carrier density and Drude weight is given by

$$D = V_F e^2/h \sqrt{\pi|n|}, \quad (7)$$

where $V_F$ is the Fermi velocity, $e$ is elementary charge, $h$ is the Planck constant, and $n$ is the carrier density at each gate voltage. Figure. 2 (d) shows the estimated carrier density and scattering rate as a function of an effective gate voltage. These values are also in good agreement with a simple theoretical prediction curve based on the parallel plate capacity model (n = $CV/e$).

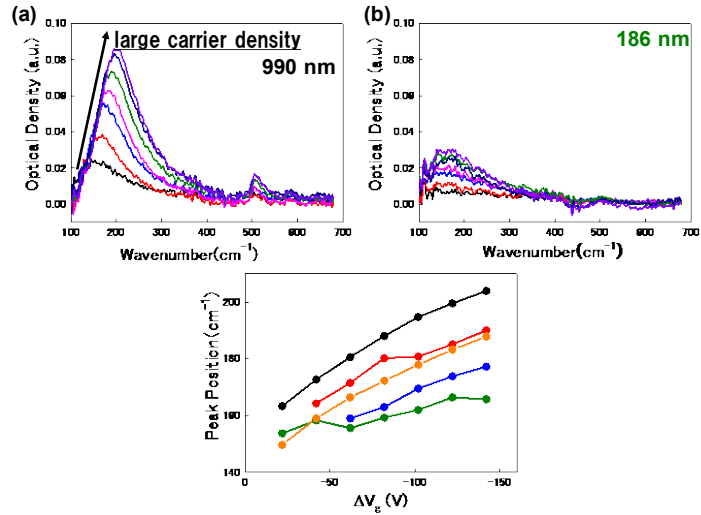

Figure 3. (color online)
(a) and (b) resonance FIR absorption spectra all graphene devices with varied widths. By applying back-gate voltage, carrier density gradually increased, as illustrated in Fig. 2. Resonance peak positions also shifted to lower frequency wavelength. (c) Peak position as function of variable gate-voltage from Dirac point estimated from Drude absorption, as illustrated in Fig. 2.

The localized plasmon resonance of the 990-nm sample is indicated in Fig. 3(a). The peak intensity significantly increased by increasing carrier density. The peak position also clearly shifted to a higher frequency depending on the increase in carrier density. For the narrowest sample as shown in Fig. 3(b) (186 nm), the peak position was more stable against changing carrier densities. To estimate the transition point of this behavior, the medium-wide samples were also tested with the same measurement and analysis procedures. Figure. 3(c) indicates the peak position as a function of the effective gate voltages in each sample. The 330- to 990-nm samples had almost the same slope dependence on the changing carrier density. However, only the narrowest sample had less dependence on gate voltage and carrier density.

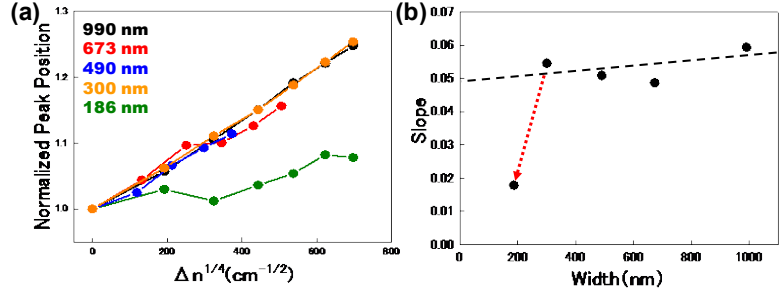

Figure 4. (color online)
(a) Normalized resonance peak position as function of estimated carrier density $\sqrt[4]{n}$ based on two-dimensional mass-less particle plasmon theory. All devices showed clear linear dependence on $\sqrt[4]{n}$. (b) The slope of each width sample. Slope of 186-nm-wide sample, showed quite small slope of carrier density.

For a more detailed comparison, the peak position is plotted as a function of $\sqrt[4]{n}$ predicted as two-dimensional plasmons in Fig. 4(a). The peak positions are normalized by smallest carrier density value and the horizontal axis is also shown as differences of carrier density from smallest position. The widest sample (990 nm) showed almost completely linear dependence on $\sqrt[4]{n}$. These results strongly indicate that the localized plasmon resonance clearly originate from the mass-less Dirac Fermions in these resonance cavities. For the 186-nm-wide sample, however, the slope of the peak shift was suddenly suppressed, and the value of the slope was less than half that of the other wider samples, as shown in Fig. 4(b). These unique behaviors are clearly explained with the model suggested by Sarma and Hwang[13]. For understanding the dimension transition, the relationship between confinement strength and Fermi wavelength is important to realize the quantized carrier motions in one dimension[18,19]. In graphene devices, the quantized conductance in a quantum wire configuration is observed in width- and carrier-density-controllable devices[20,21]. For the narrowest sample in one of these studies (230 nm), the quantized conductance was observed in a wide carrier density range from the CNP to $1.5 \times 10^{12} cm^{-2}$ at 4.2K[21]. In our narrowest device (186 nm), the carrier density modulated from $2.6 \times 10^{11} cm^{-2}$ to $1.0 \times 10^{13} cm^{-2}$. The quantized conductance might be less than ten channels with this device under the lowest carrier density. These estimations are consistent with our dimensionality transition observations. Similar behavior was recently observed in carbon nanotube (CNT) plasmon resonances[22–24]. These CNTs had strongly confined one-dimensional carriers due to naturally rolled-up structures of less than 10 nm in diameter. Therefore, these CNT plasmons are robust against chemical doping, which is stronger than electrical back-gate control of carrier densities. In the case of CNTs, however, it is difficult to precisely control both confinement strength and carrier density. In graphene devices, the plasmon behavior is completely controllable by changing the sample dimensions and external electric field such as the gating effect. Therefore, this result is important for more controllable and higher efficient optical devices based on localized plasmon resonance.

**Conclusion**

We first demonstrated the possibility of dimensionality transition behavior of graphene localized plasmon resonances from two dimensional plasmon to a one-dimensional one. From a detailed analysis based on carrier density, which is estimated from the Drude absorption behavior on the same wafer and graphene sheet, the wide devices showed $\sqrt[4]{n}$ dependence on the peak shift of plasmon resonance. For the narrowest sample (186 nm), the carrier-density dependence was strongly suppressed. These behaviors are consistent with theoretical prediction based on the mass-less Dirac Fermion in the linear dispersion relationship. These results also indicate that we can control the carrier-density dependence and robustness against environment effects. This is important to fabricate highly efficient optical devices based on low-dimensional plasmons.

**Acknowledgments**

This work was supported by JSPS KAKENHI Grant Number JP25286018.

**Figure Captions**

Figure 1. (color online)
(a) Schematic of measurement setup. FIR spectrum was measured at patterned area, and electric transport measurement was conducted at non-lithographic area. Gate-voltages were applied using back-gate configurations. (b) Optical image of sample holder used in this study. Samples were mounted on FT-IR sample holder. Electric contacts were done using hand-crafted probing systems. (c) Schematic of sample pattern and AFM images obtained after developed devices. Cavities of optical resonances were same length ($L$) but with variable widths ($W$) and connected to carrier injection lines.

Figure 2. (color online)
(a) Source-drain current as function of back-gate voltage. In air, Dirac point was not observed due to shift by molecular adsorption (red curve). Blue curves indicate transport properties under vacuum. (b) Drude absorption in FIR region observed with non-patterned region at each gate voltage. By increasing back-gate voltages, Drude absorption monotonically increased by increasing carrier densities. (c) Carrier densities and scattering-time constant estimated from fitting of Drude absorptions. Carrier densities increased by applying gate voltages, and scattering-time constant was almost constant due to electrically carrier modulations.

Figure 3. (color online)
(a) and (b) resonance FIR absorption spectra all graphene devices with varied widths. By applying back-gate voltage, carrier density gradually increased, as illustrated in Fig. 2. Resonance peak positions also shifted to lower frequency wavelength. (c) Peak position as function of variable gate-voltage from Dirac point estimated from Drude absorption, as illustrated in Fig. 2.

Figure 4. (color online)
(a) Normalized resonance peak position as function of estimated carrier density $\sqrt[4]{n}$ based on two-dimensional mass-less particle plasmon theory. All devices showed clear linear dependence on $\sqrt[4]{n}$. (b) The slope of each width sample. Slope of 186-nm-wide sample, showed quite small slope of carrier density.

**Figures**

Figure 1.

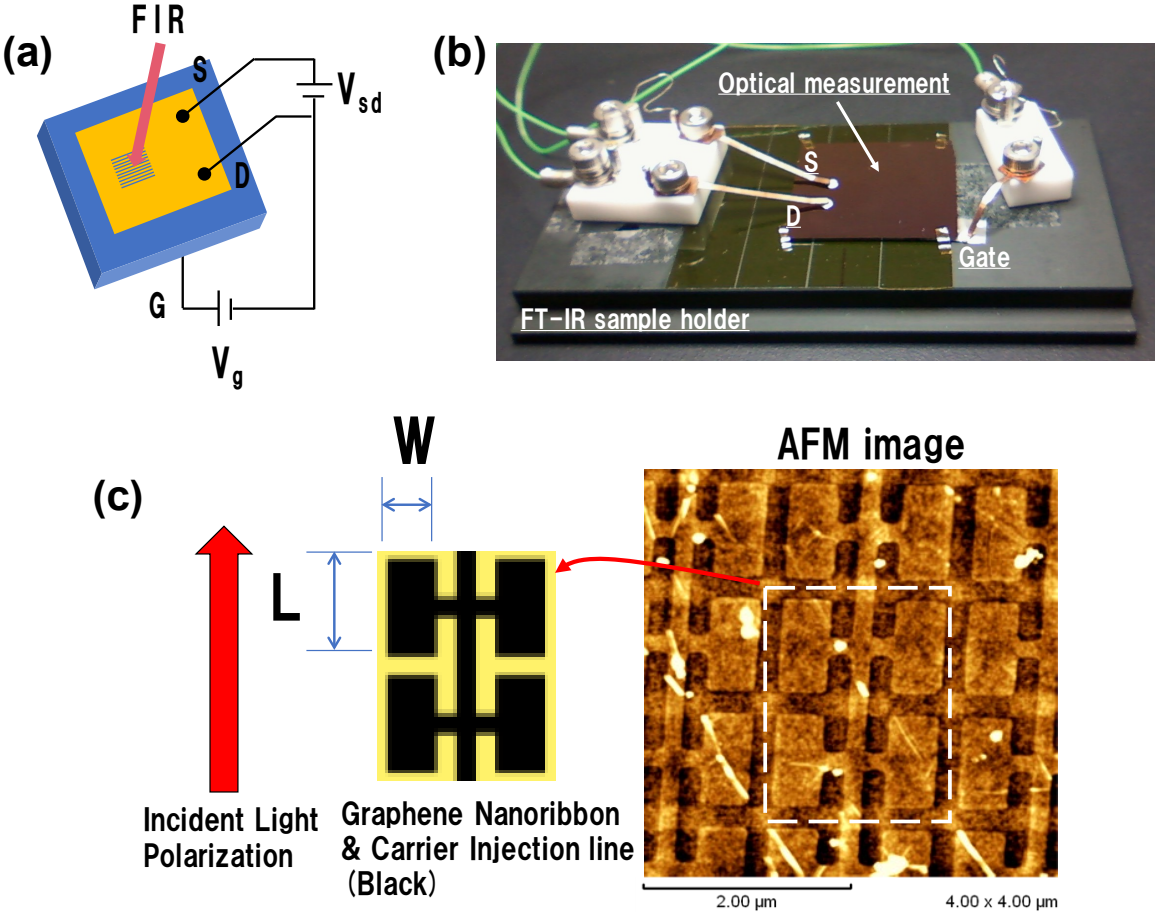

Figure 2.

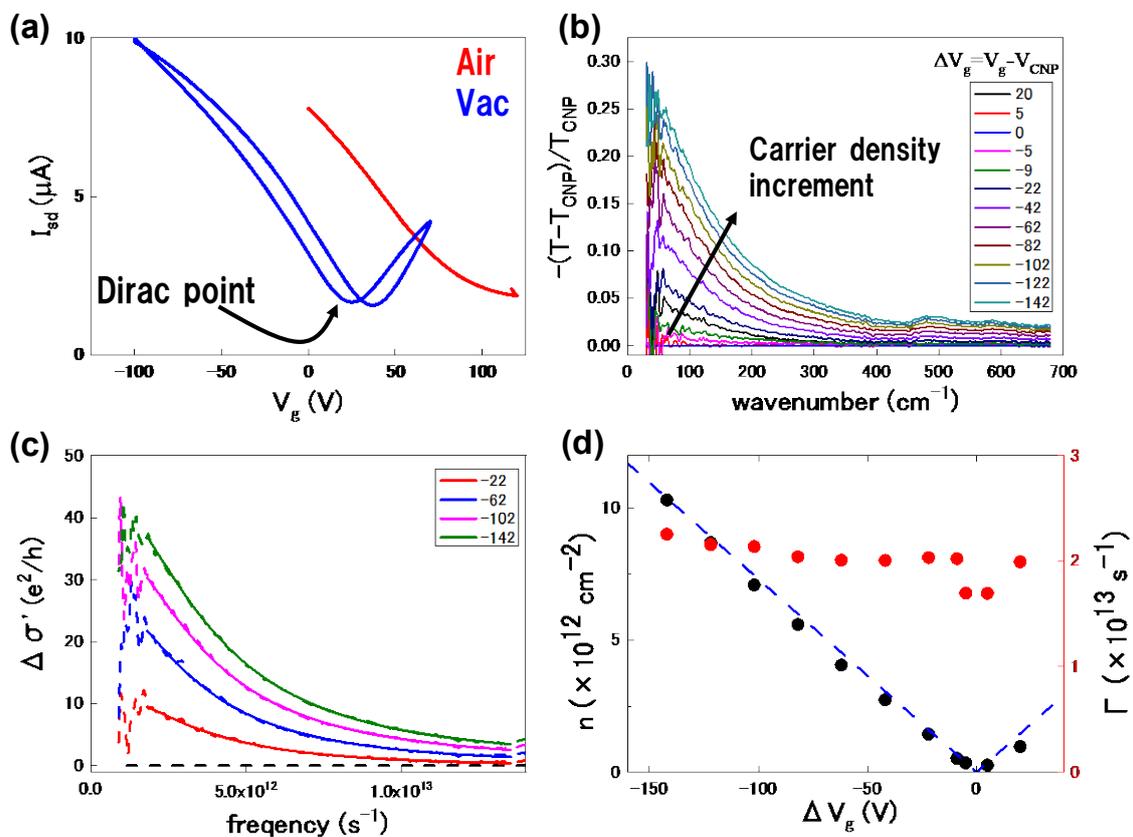

Figure 3.

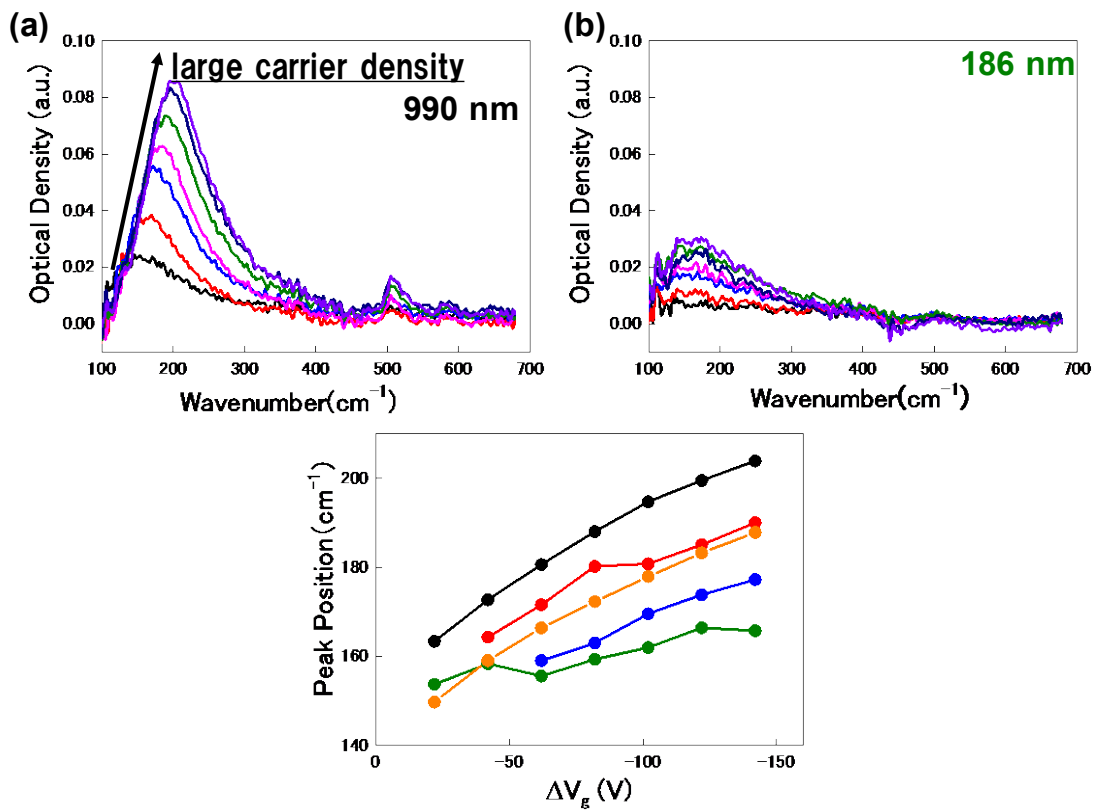

Figure 4.

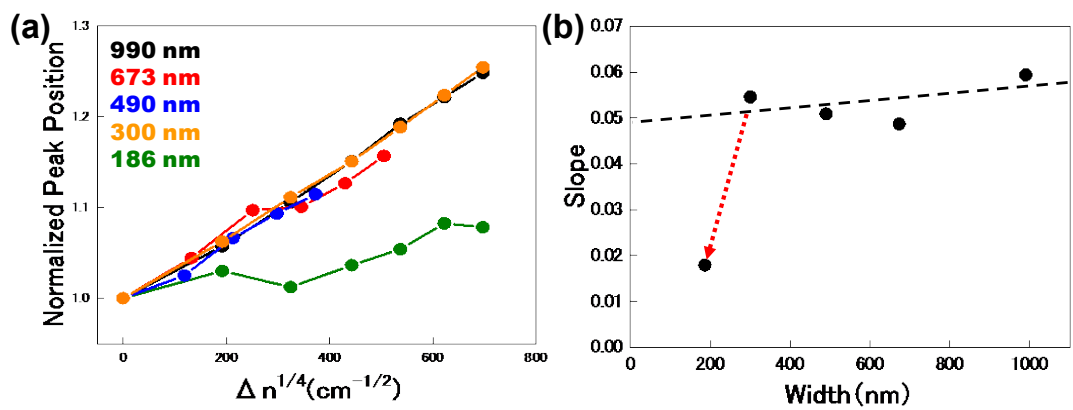